# The Faulkes Telescope Project: Not Just Pretty Pictures

F. Lewis, P. Roche

Las Cumbres Observatory Global Telescope Network (LCOGTN), Faulkes Telescope Project, Open University

**Summary** The Faulkes Telescope (FT) Project is an educational and research arm of the Las Cumbres Observatory Global Telescope Network (LCOGTN). As well as producing spectacular images of galaxies, nebulae, supernovae remnants, star clusters, etc, the FT team is involved in several projects pursuing scientific goals. Many of these projects also incorporate data collected and analysed by schools and amateur astronomers.

We detail a few of these projects in forthcoming sections.

## 1 Introduction

The Faulkes Telescope Project currently has two 2-metre robotic telescopes, located at Haleakala on Maui (FT North) and Siding Spring in Australia (FT South). It is planned for these telescopes to be complemented by a research network of eighteen 1-metre telescopes, along with an educational network of twenty-eight 0.4-metre telescopes, providing 24 hour coverage of both northern and southern hemispheres.

## 2 Observations of Occultations of Uranian moons in conjunction with Apostolos Christou (Armagh Observatory)

The 2007 Uranian Equinox allowed unique observations of the planet, its rings and satellites, possible only twice during the planet's 84 year orbit. Among these were mutual eclipses and occultations between the 5 classical satellites: Ariel, Umbriel, Titania, Oberon and Miranda.

Several of these events were observed using both FT North and FT South. To mitigate against Uranus' glare, the SDSS-i' band filter was used.

Six positive detections were made (three eclipses and three occultations), starting with an occultation of Umbriel by Oberon on $4^{th}$ May, 2007[1] (see Fig. 1) and concluding with an eclipse of Miranda by Ariel on $30^{th}$ November 2007.

Preliminary analysis of the data indicates the existence of small yet statistically significant offsets between the observed and predicted positions of the satellites, of order a few hundred km at Uranus or a few tens of milliarcsec on the sky plane. We anticipate that this new data will yield improved estimates of the satellite masses and Uranian gravity harmonics.

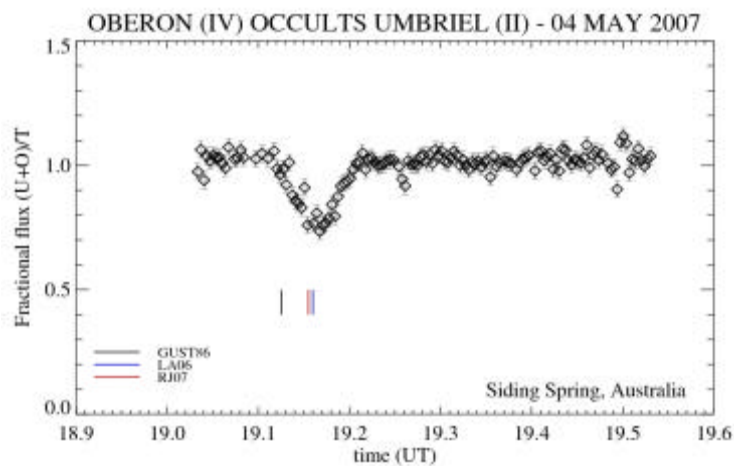

**Fig. 1** Lightcurve of Oberon's occultation of Umbriel, May $4^{th}$, 2007

## 3 The Imaging of a Fast Rotating Asteroid by Richard Miles (British Astronomical Association)

Long exposures of asteroids often produce a 'streak' across the image, rather than a point source, as the asteroid moves rapidly through the field-of-view. On occasion, this streak (Fig. 2) appears to have a banded structure, which can denote that the asteroid is rotating rapidly, and that the brightness of the object changes as different parts of its surface face the Earth.

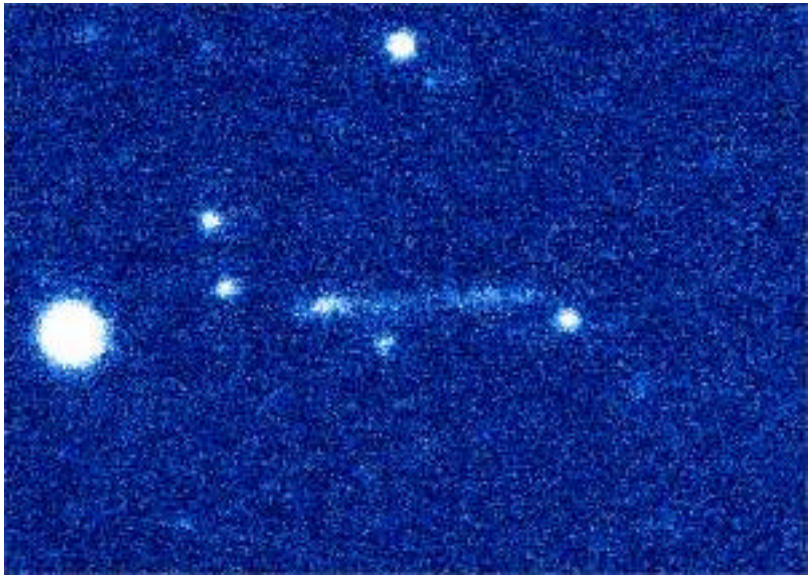

**Fig. 2** FT South image of the asteroid 2008 HJ (streak in centre, left to right)

Using FT South, amateur astronomer Richard Miles, in conjunction with several schools, showed that the asteroid 2008 HJ completes a full rotation every 42.7 seconds (Fig. 3), making it the fastest rotating natural object known in our Solar System. This discovery was subsequently ratified by the International Astronomical Union on 22nd May, 2008. The previous record holder was the asteroid 2000 DO8, which rotates every 78 seconds.

This latest discovery was the most recent outcome of a new project to survey the properties of small (<150-metre) near-Earth asteroids. The project had an early success in April 2008, having found that asteroid 2008 GP3 rotates once every 11.8 minutes. Asteroid 2008 HJ was only the fourth object observed as part of this study.

The observations suggest that 2008 HJ is a compact stony object some 12m x 24m in size, smaller than a tennis court yet probably having a mass in excess of 5,000 tonnes. It was moving at almost 45 kilometres per second (more than 100,000 mph) when it hurtled past the Earth in late April.

It appears that the chances of finding similar objects are high and a challenge has been set to FT users to find objects spinning even faster than 2008 HJ. Cooperation between all the observing groups, whether they are astronomers or schools students, will be essential if asteroid rotation rates are to be accurately identified.

Our knowledge of the near-Earth population of small asteroids is very sparse, so FT users are contributing directly to our understanding of these nearest neighbours of ours. It is believed that most of these objects are probably fragments ejected from collisions between larger bodies which took place some time in the distant past. However, other objects may have originated when the solar nebula was formed over 4.6 billion years ago.

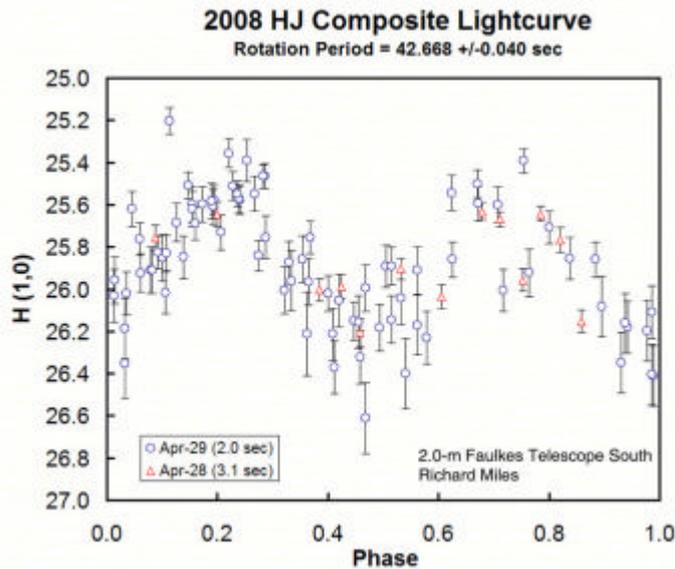

**Fig. 3** Lightcurve of the asteroid, 2008 HJ, April, 2008

## 4 The YORP Effect as studied by Alan Fitzsimmons, Stephen Lowry and colleagues (Queens University, Belfast)

Observations from FT formed a part of a dataset to demonstrate the existence of the theoretical 'YORP effect' for the first time. By using a range of telescopes, including FT North, it was discovered that the asteroid, (54509) 2000 PH5, is rotating faster by 1 millisecond every year.

The Yarkovsky-O'Keefe-Radzievskii-Paddack (YORP) effect is believed to alter the way small bodies in the Solar System rotate. The effect occurs when photons from the Sun bounce off the irregular surface of the asteroid generating a force (no such force would be generated on an ellipsoid with a totally smooth surface). By analogy, if one were to shine light on a propeller over a long enough period, it would start spinning.

Although this is an almost immeasurably weak force, astronomers believe it may be responsible for spinning some asteroids up so fast that they break apart, perhaps leading to the formation of binary asteroids. Others may be slowed down so that they take many days to rotate once. The YORP effect also plays an important role in changing the orbits of asteroids between Mars and Jupiter, including their delivery to planet-crossing orbits. Despite its importance, the effect has never been seen acting on a Solar System body, until now.

Shortly after its discovery in 2000, it was realized that this asteroid would be the ideal candidate for such a YORP detection. At just 114m in diameter, it is relatively small and so more susceptible to the effect. Also, it rotates very fast, with one day on the asteroid lasting just over 12 Earth minutes, implying that the YORP effect may have been acting on it for some time. With this in mind, the team of radar and optical astronomers undertook a long term monitoring campaign of the asteroid with the aim of detecting any tiny changes in the spin-rate.

Four years of observations were made and, during the same time period, a team led by Patrick Taylor and Jean-Luc Margot of Cornell University used the Arecibo Observatory in Puerto Rico and the Goldstone facility in California to observe the asteroid using radar[2]. They were able to reconstruct a 3-D model of the asteroid's shape (Fig. 4), with the necessary detail to allow a theoretical YORP value to be calculated.

With careful analysis of the optical data, the asteroid's spin rate was seen to steadily increase with time, at a rate that can be explained by YORP theory. Critically, the effect was observed year after year. Furthermore, this result was elegantly supported via analysis of the combined radar and optical data, as it was required that the asteroid be increasing its spin-rate at exactly this rate in order for a satisfactory 3-D shape model to be determined.

Lowry et al.[3] performed detailed computer simulations and found that the asteroid's orbit about the Sun could remain stable for up to 35 million years in the future, allowing the spin-rate to be reduced to just 20 seconds, faster than any asteroid spin-rate ever seen. This exceptionally fast spin-rate could force the asteroid to reshape itself or even split apart, leading to the birth of a new binary system.

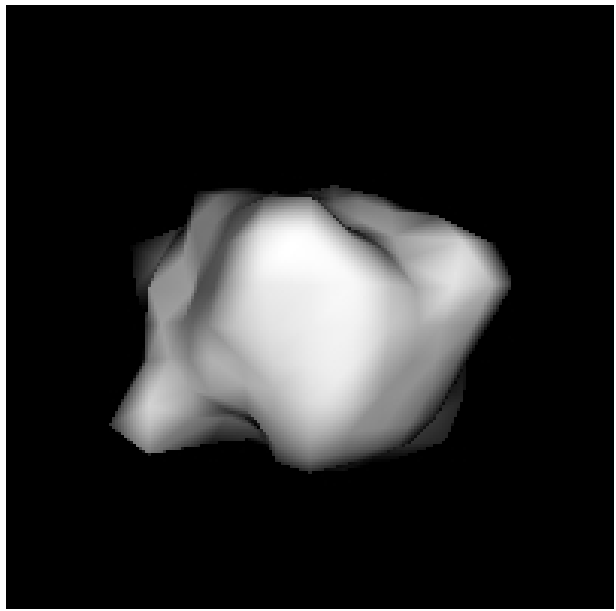

**Fig. 4** 3D view of the asteroid, 2000 PH5

## 5 Monitoring LMXBs with the Faulkes Telescopes by Fraser Lewis and David Russell (University of Amsterdam)

We have been undertaking a monitoring project of 13 Low-Mass X-ray binaries (LMXBs) using FT North since early 2006. The introduction of FT South has allowed us to extend this to monitor a total of 30 LMXBs. With new instrumentation, we also intend to expand this monitoring to include both infrared wavelengths (z and y band) and spectroscopy.

The aims of the project are:

1. To identify transient outbursts in LMXBs. LMXBs may brighten in the optical/near-infrared for up to a month before X-ray detection. The behaviour of the optical rise is poorly understood, especially for black hole X-ray binaries. Catching outbursts from quiescence will allow us to examine this behaviour and alert the astronomical community to initiate multi-wavelength follow-up observations.

2. To study the variability in quiescence. Recent results have suggested that many processes may contribute to the quiescent optical emission, including emission from the jets in black hole systems[4]. By monitoring the long-term variability of quiescent LMXBs, we will be able to provide constraints on the emission processes and the mass functions.

Preliminary studies suggest that our sources split into three categories

Those that show little or no variability in quiescence

Those that show some non-periodic variability in quiescence

Those that undergo outbursts

One such outbursting source, GX 339-4, is detailed below

### 5.1 GX339-4

A target for FT South is the transient black hole binary GX 339-4, which went into outburst in early 2007, followed by a steady decline in the following months (Fig.5). The outburst was detected at X-ray[5], optical/IR[6] and radio[7] wavelengths. Our observations show that the source continues to decline in V, R and i' bands until ~ MJD 54585 when the source increased in brightness to its previous brightest level[8].

Our ATel has triggered multi-wavelength follow-ups with the South African Large Telescope (SALT), the Very Large Telescope (VLT) and the Swift gamma-ray and RXTE X-ray satellites.

We also note rapid flaring in i' and V band with jumps of ~ 0.3 – 0.4 magnitudes in periods of ~ 140 seconds.

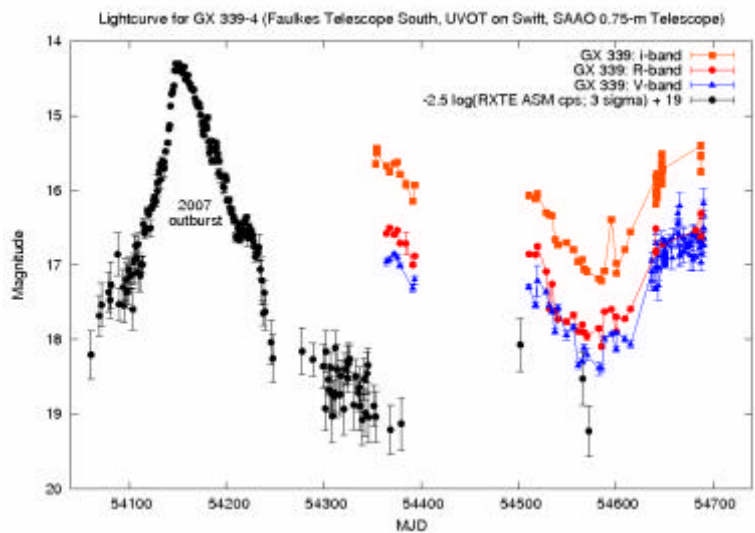

**Fig. 5** FT multi-wavelength data, combined with RXTE X-ray observations showing the system's 2007 and 2008 outbursts

## 6 Exoplanet studies conducted by Rachel Street, Tim Lister, Marty Hidas, Nairn Baliber and others, LCOGTN

LCOGTN researchers are involved in the Wide Angle Search for Planets (SuperWASP) consortium, primarily in the follow-up work of exoplanet candidates detected by microlensing and transit surveys. Their work comprises two methods of detection, microlensing and transiting sources.

The technique used by SuperWASP involves two sets of cameras (in the Canary Islands and South Africa) to watch for events known as transits, where a planet passes directly in front of a star and blocks out some of the star's light. From the Earth the star temporarily appears a little fainter.

The SuperWASP cameras work as robots, surveying a large area of the sky at once. Each night astronomers receive data from hundreds of thousands of stars. They can then check for transits and hence planets. The transit technique also allows scientists to deduce the size and mass of each planet.

One such exoplanet, WASP-10[9] has been studied and a lightcurve of a transit event is shown (Fig. 6).

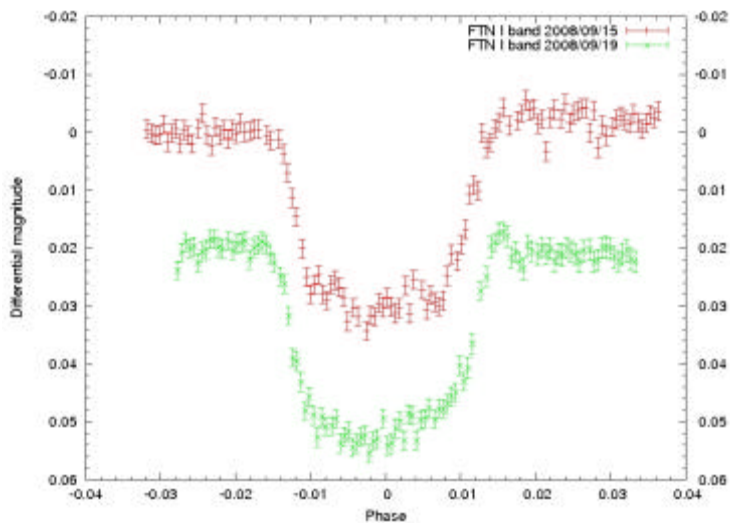

Fig 6. Lightcurve of transit of WASP-10 in I band

## 7 Acknowledgements

FL acknowledges support from the Dill Faulkes Educational Trust. We thank the contributing scientists for their data.